      [1999/12/01 v1.4c Il Nuovo Cimento]
\documentclass{cimento}

\newcommand{\eh}[1]{\,\mathrm{#1}}

\newcommand{\ttt}[1]{\times10^{#1}}

\newcommand{\dg}{^{\circ}}

\newcommand{\pcnt}[1]{$#1\eh{\%}$}

\renewcommand{\epsilon}{\varepsilon}

\usepackage{graphicx}  
\title{The MAGIC Telescopes - Status and Recent Results}
\author{S.~Klepser\from{ins:x} for the MAGIC collaboration}
\instlist{\inst{ins:x} IFAE, Edifici Cn., Campus UAB, E-08193 Bellaterra, Spain}
\begin{document}

\maketitle

\begin{abstract}
The MAGIC telescopes are two Imaging Atmospheric Cherenkov Telescopes
located on the canary island of La Palma. They provide the lowest
energy threshold among existing instruments of the kind, reaching down to 50 GeV in standard
trigger mode. This allows us to close the energy gap between
satellite-bourne and ground-based gamma-ray observations for strong enough
sources. During the first five years of monoscopic observations, many interesting results could thus be achieved.
With the second MAGIC telescope, which started operation in 2009, the
sensitivity could be improved by stereoscopic imaging, and 5 new
detections could already be reported in 2010. We present the status of the MAGIC
telescopes in 2010 and review the latest results obtained in mono- and
stereoscopic
mode. This includes, among others, the detection of the head-tail galaxy IC~310, a new
multiwavelength study of Mrk~501, an updated lightcurve of the Crab Pulsar.
\end{abstract}

\section{Introduction}

The two MAGIC telescopes are presently one of the leading facilities for the
observation of very-high energy (VHE, above hundreds of
GeV) gamma rays. They indirectly detect the gamma-rays by recording images of
gamma-ray initialized particle cascades in the upper atmosphere, the so-called Cherenkov
imaging technique. The location of the instrument is in the Roque de los
Muchachos Observatory on the
the Canary Island of La Palma ($28.8\dg$ N, $17.8\dg$ W), at an altitude of
2220 m a.s.l.

Among the few other instruments of that kind, it is
the one with the lowest energy threshold, reaching down to about
$50\eh{GeV}$, while still being sensitive up to several tens of TeV. This low
energy threshold is a merit of several factors, among which the huge mirror
dishes of $17\eh{m}$ diameter, the fast ($2\eh{GHz}$) readout
electronics and the altitude are the most prominent ones.
Besides the lowest threshold, MAGIC is built using a light-weight carbon fibre
telescope frame. This makes it the fastest telescope concerning
repositioning to a given direction in the sky. It allows for an $180\dg$
turn within about $20-40\eh{s}$, which is essential for observations of gamma ray
bursts (GRBs).

The operation of MAGIC started in fall 2004, using only the MAGIC-I telescope
(MAGIC Phase I). It was complemented by a second telescope which became
fully operational in late 2009 (Phase II). The stereoscopic observation of the
particle showers improves the performance of the system, and lowers the
systematic uncertainties.
As a benefit of our improved performance, 5 new extragalactic VHE sources could
already be reported in 2010\footnote{http://www.astronomerstelegram.org}.

\section{Performance}

A detailed paper about the MAGIC performance in stereoscopic mode
is in preparation. The following numbers may hence still be
subject to minor changes and should be considered preliminary. A detailed
description of the monoscopic performance is given in~\cite{ref:magcrab}.

\begin{figure}
\centering
\includegraphics[width=0.55\columnwidth]{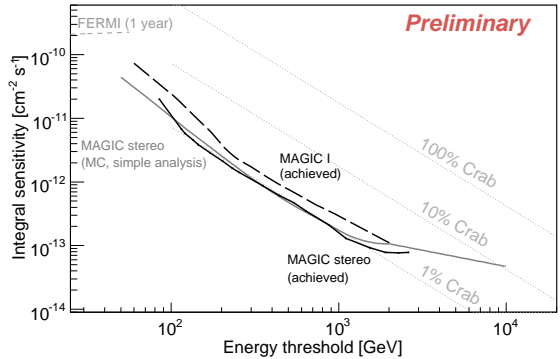}     
\caption{MAGIC integral sensitivity in Phase I (mono) and II (stereo),
estimated from $3.5\eh{h}$ of Crab Nebula data. The
achieved sensitivity for Phase II are
better then the prediction mainly because they exploit fully optimized
reconstruction algorithms that were not available at the time the MC was
produced.}
\label{fig:sens}
\end{figure}

Figure \ref{fig:sens} shows the sensitive range of monoscopic Phase I
observations, and the predicted and achieved sensitivity for the stereoscopic
Phase II. Here, the sensitivity is expressed as the source flux that is
detectable with $5\eh{\sigma}$ confidence within 50 hours of data taking at
low zenith angle. The expected sensitivity was derived from a Monte Carlo
simulation (MC). From Phase I to Phase II, it improved from \pcnt{1.6} to
less than \pcnt{1} of the Crab Nebula flux, which means up to \pcnt{60} less
required observation time for the same detection.

The energy resolution above $300\eh{GeV}$, defined as RMS, improved from
\pcnt{25} in phase I to \pcnt{15} in phase II. The \pcnt{39} directional resolution (i.e. Sigma of a
2D Gaussian function) at these energies could be improved from $0.1\dg$ to
$<0.07\dg$. This, along with the better gamma/hadron separation in
stereoscopic mode, reduces the background by about a factor of 3, which leads
to the above improvement in sensitivity and also leads to smaller systematic
uncertainties, especially at low energies. It allows us to image the Crab
Nebula even at energies around $80\eh{GeV}$ (see Figure~\ref{fig:crabLE_ic310}),
where the angular resolution is still about $0.15\dg$.


\section{Scientific Scope and Selected Recent Results}

MAGIC has a wide field of scientific targets, only few of which can be briefly presented in
these proceedings. The topics that cannot be discussed in the following review
include for example the flux upper limits derived for 
gamma-ray bursts (see e.g.~\cite{ref:grb}), possible dark matter annihilation
spots in dwarf spheroidals~\cite{ref:wilman} and
galaxy clusters~\cite{ref:perseus}, and the
globular cluster M13, which yielded interesting constraints on the density
and/or efficiency of its pulsar population \cite{ref:m13}. A complete list of publications can be found
on the MAGIC website\footnote{http://wwwmagic.mppmu.mpg.de}.

\subsection{Extragalactic Observations}

\begin{figure}
\centering
\includegraphics[width=0.40\columnwidth]{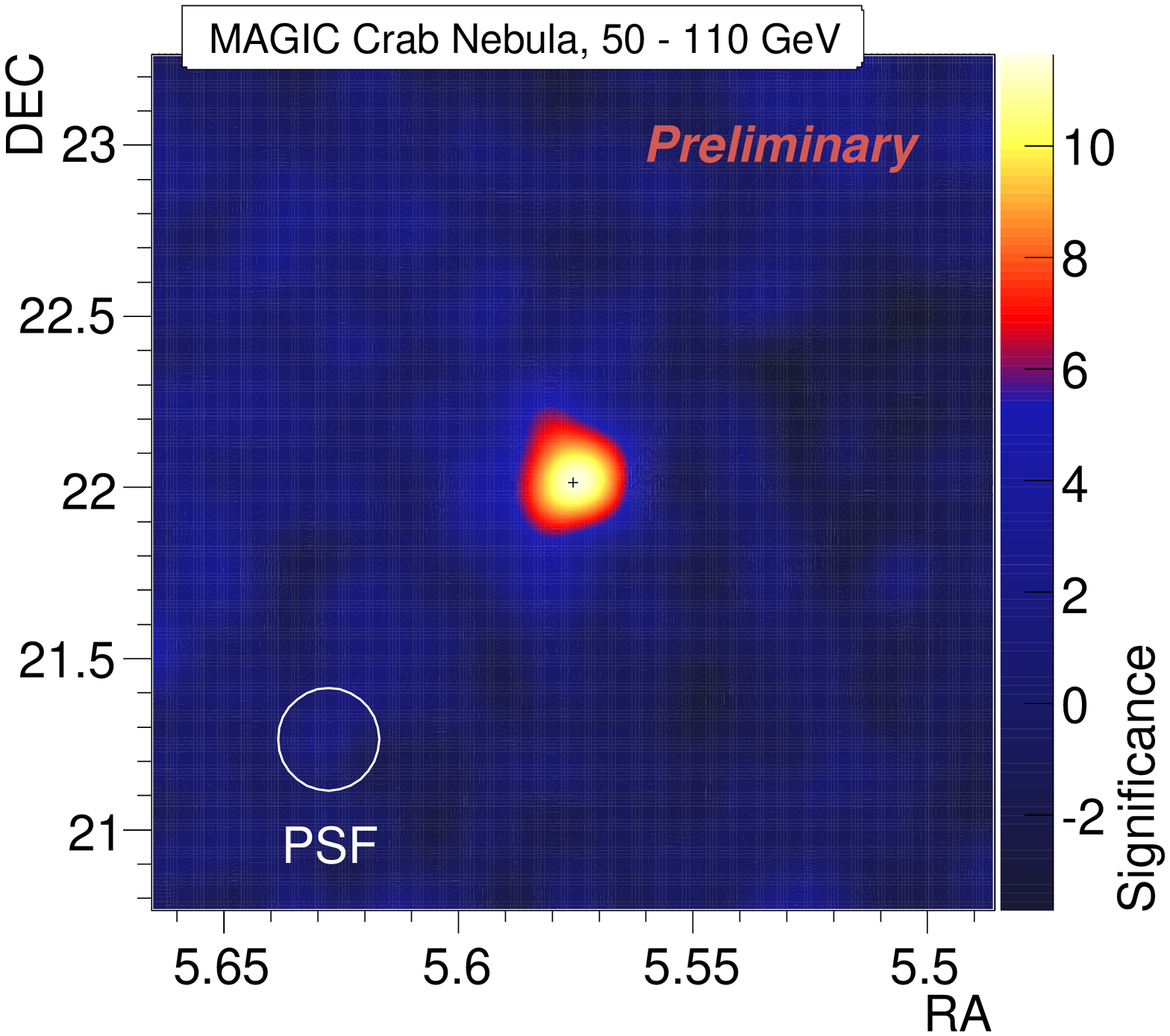}     
\includegraphics[width=0.50\columnwidth]{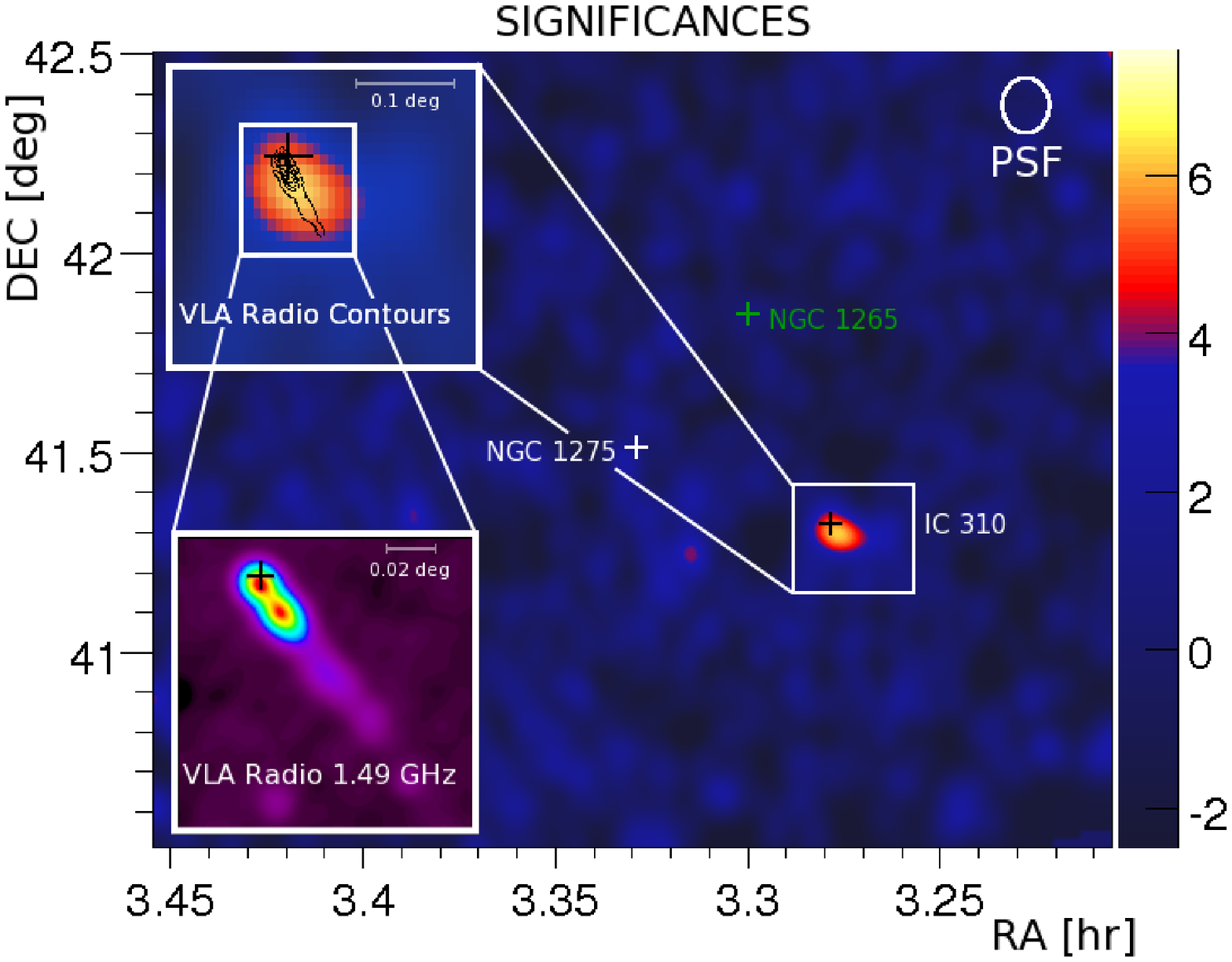}     
\caption{Left: Low energy skymap of $2.6\eh{h}$ of Crab Nebula data below $30\dg$ zenith angle.
The plot was done from events with reconstructed energies below $100\eh{GeV}$,
leading to a range of true energies of about $50$ - $110\eh{GeV}$ (Median
$80\eh{GeV}$). The PSF circle shows the \pcnt{39} resolution, see text. Right: Significance skymap of the IC~310 region for energies above
$400\eh{GeV}$, in comparison to radio contours from NVSS
(see~\cite{ref:ic310} for a detailed discussion and references).}
\label{fig:crabLE_ic310}
\end{figure}

Being on the northern hemisphere, MAGIC is ideally located to observe
the gamma ray emission of extragalactic objects. Most of these are active
galactic nuclei (AGN) of different kinds.
One of the first sources MAGIC discovered in stereoscopic mode is the head-tail radio galaxy IC~310~\cite{ref:ic310}
(Figure~\ref{fig:crabLE_ic310}).
It was detected in $20.6\eh{h}$ of observations
taken between 2009 October and 2010 February.
The observed spectral
energy distribution (SED) is remarkably flat, (photon index $\gamma=-2.00 \pm
0.14$),
and there are clear hints for short-scale variability. This result
favours the VHE emission to originate from
the inner jet, close to the central engine of the AGN. The wide, flat spectrum
furthermore challenges simple synchrotron self-compton (SSC) models, favouring
more complex or hadronic models for the emission mechanism.

\begin{figure}
\centering
\includegraphics[width=0.65\columnwidth]{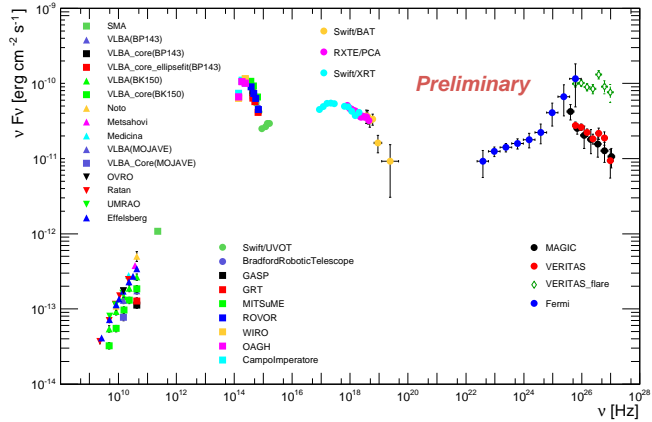}     
\caption{Spectral energy distribution of Mrk~501 averaged over all
observations taken during the MWL campaign performed between 2009 March 15 and
2009 August 1. The TeV data from MAGIC and VERITAS have been corrected for the
absorption in the extragalactic background light (find details on this, and
further references, in~\cite{ref:mrk501}).}
\label{fig:mrk501}
\end{figure}

For strong, established AGNs, MAGIC is frequently participating
in large, synchronized multiwavelength (MWL) campaigns, organized on long terms
between many observatories worldwide. Figure~\ref{fig:mrk501} gives an example
SED from a MWL campaign on Mrk~501, taken during 4.5 months in
2009~\cite{ref:mrk501}. This
radio to TeV campaign provided the most detailed SED yet collected for that
source. A multi-frequency SED allows for a deeper understanding of a source -
in this case, the SED is well described by a one-zone SSC model, with an
emission region $\le 0.1\eh{pc}$ and an electron population with two spectral
breaks. 

Two complementary topics to the above are the study of the extragalactic
background light (EBL) and the extragalactic magnetic field (EGMF). The
EBL, mainly comprising diffuse light from stars and dust, leads to the
absorption of gamma rays above few hundreds of GeVs. The sub-$100\eh{GeV}$
threshold of MAGIC however allowed us to detect the most distant VHE
AGN~\cite{ref:ebl}. With few basic
assumptions about plausible intrinsic emission spectra, we
could exclude several EBL modelings, showing that the universe is more
transparent than was previously assumed.

Testing the emission from AGNs for a possible extended halo was recently shown to probe the
strength of the EGMF \cite{ref:halo}. As described there, VHE gamma rays from
AGNs may undergo interactions in the extragalactic space between the AGN and
ourselves that produce intermediate charged particles and secondary gamma rays. If
the EGMF has a strength of few $10^{-15}\eh{G}$, the charged particles may
both live long enough and be deflected enough to mimick a gamma-ray halo
around the otherwise point-like source.
The fact that such a
halo was not found beyond uncertainties suggests that the strength of the EGMF
must either be well above or below the value of few times $10^{-15}\eh{G}$.

\subsection{Galactic Observations}

\begin{figure}
\centering
\includegraphics[width=0.49\columnwidth]{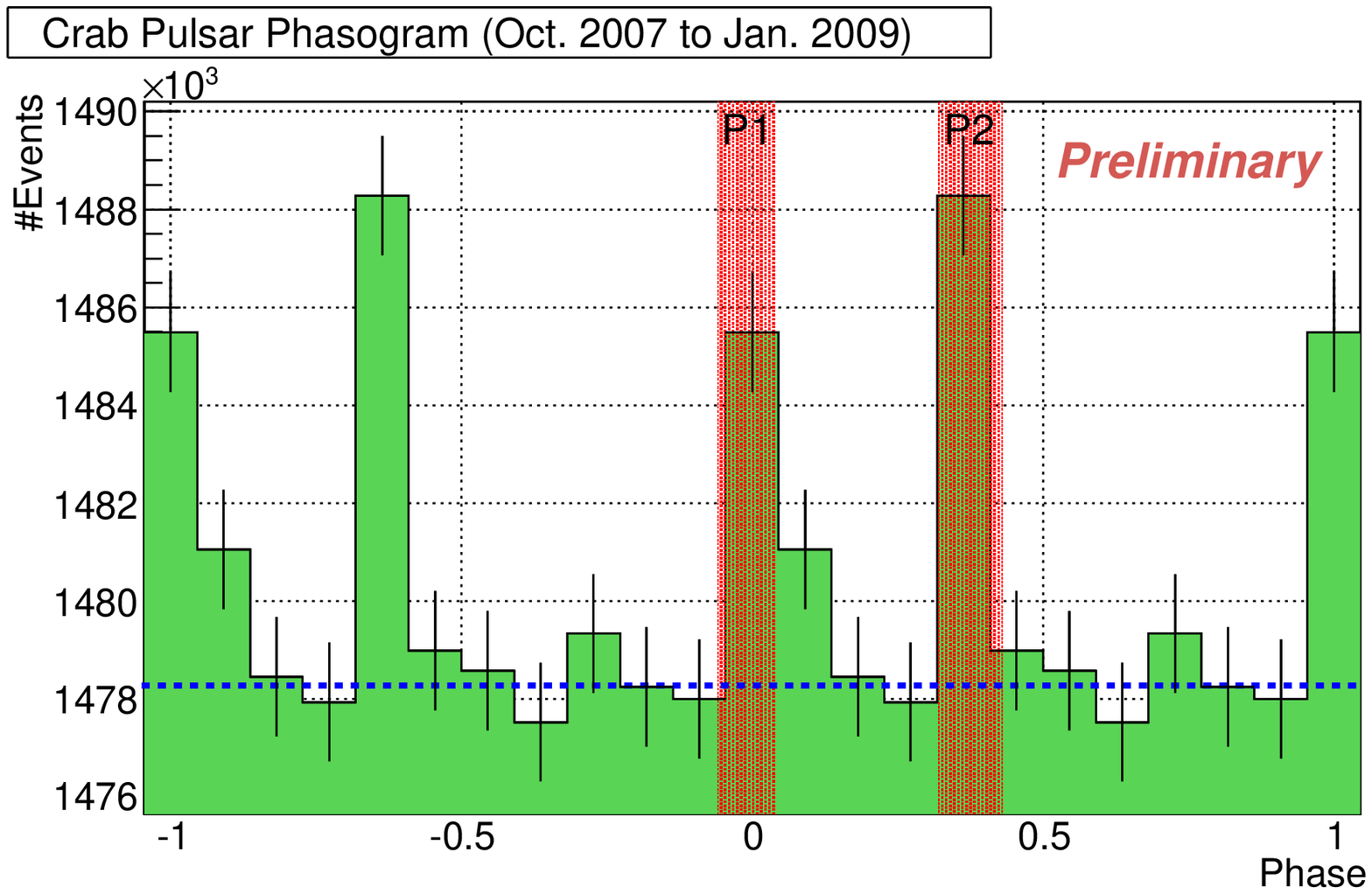}     
\includegraphics[width=0.38\columnwidth]{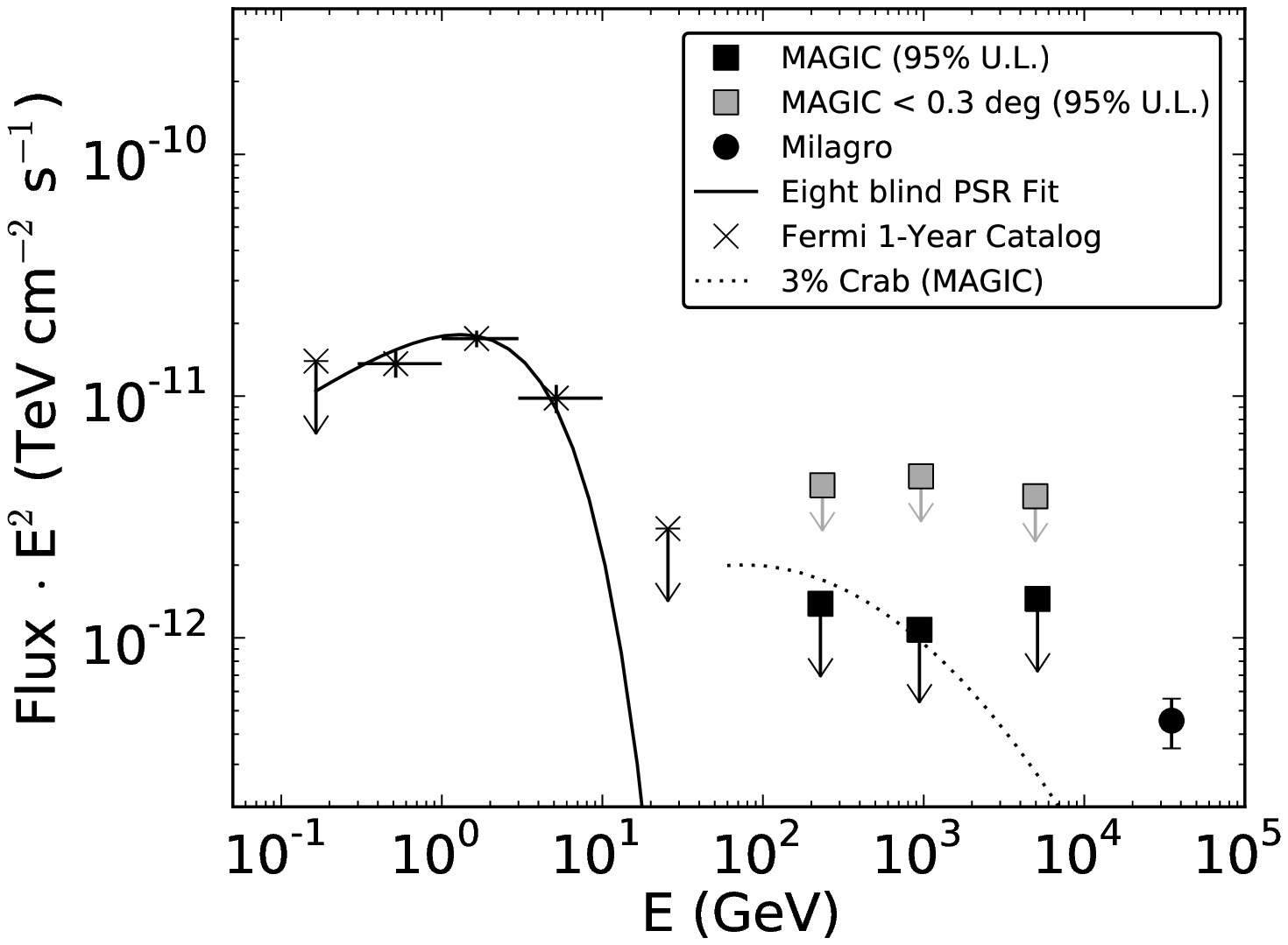}     
\caption{Left: Phaseogram of the Crab pulsar, obtained from $59.1\eh{h}$ of good
quality data, using only Cherenkov images with and integral size between 25
and 500 photo electron equivalents, roughly corresponding to energies above
$25\eh{GeV}$. Right: Differential flux upper limits for the region around
1FGL~J1954.3+2836, in the context of the Fermi and Milagro flux estimations (see~\cite{ref:g65} and references therein).}
\label{fig:crabpulsar_g65}
\end{figure}

Galactic targets of observations of MAGIC include pulsars, pulsar wind nebulae
(PWN), supernova shells, binary systems and magnetars. Due to its low
threshold, MAGIC-I is the only Cherenkov telescope to date that detected
pulsed gamma ray signals from a pulsar. The mere fact of detecting pulsed
emission above $25\eh{GeV}$ from the Crab pulsar~\cite{ref:crabpulsar} already ruled out
emission models in which the pulsed component is produced anywhere near the
pulsar surface, because the strong magnetic field would simply absorb such
highly energetic radiation by invoking electron pair production.
Figure~\ref{fig:crabpulsar_g65} shows an updated light curve from $59.1\eh{h}$ of
data collected between 2006 and 2009. The excess events of the two 
peaks are $(6.2\pm1.4)\ttt{3}$ for P1 and $(11.3\pm1.5)\ttt{3}$ for P2,
corresponding to $4.3\eh{\sigma}$ and $7.4\eh{\sigma}$, respectively.

In 2009, we took 25.5 hours of good quality monoscopic data in the region of
supernova remnant G65.1+0.6. The area hosts the two GeV sources
1FGL~J1954.3+2836 and 1FGL~J1958.6+2845, both of which were reported to have a
multi-TeV counterpart detected in Milagro data \cite{ref:milagro}. From our
observations, we could extract upper limits on the \pcnt{2-3} Crab Nebula flux
level~\cite{ref:g65} (Figure~\ref{fig:crabpulsar_g65}). This supports the idea that the Milagro emission emerges from old
and/or high-peaked PWN, similar to what was found for several
other Milagro-detected Fermi bright sources. Other recent MAGIC-I upper limits
of PWNe can be found in~\cite{ref:pwn}.

\begin{figure}
\centering
\includegraphics[width=0.43\columnwidth]{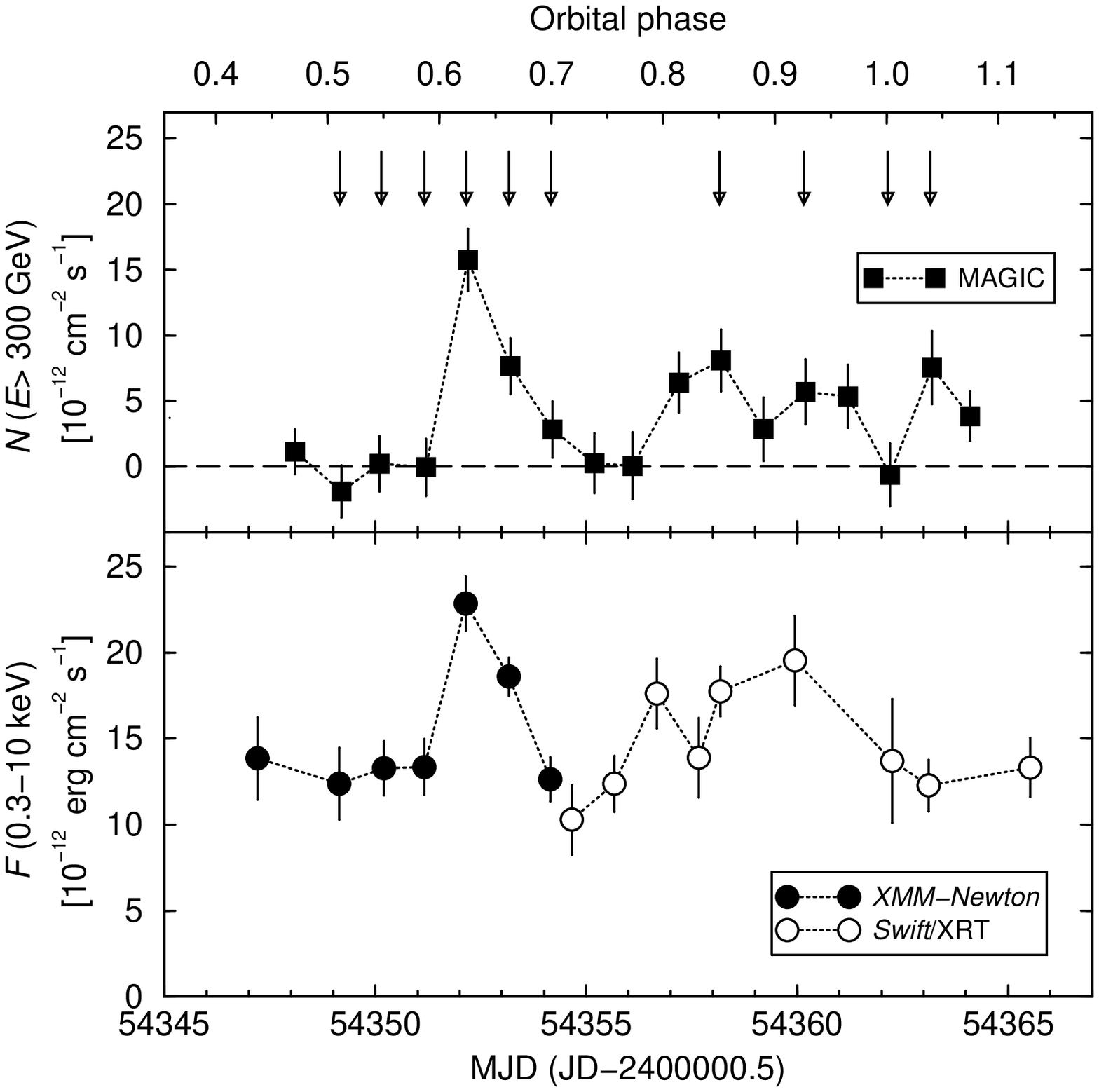}     
\includegraphics[width=0.45\columnwidth]{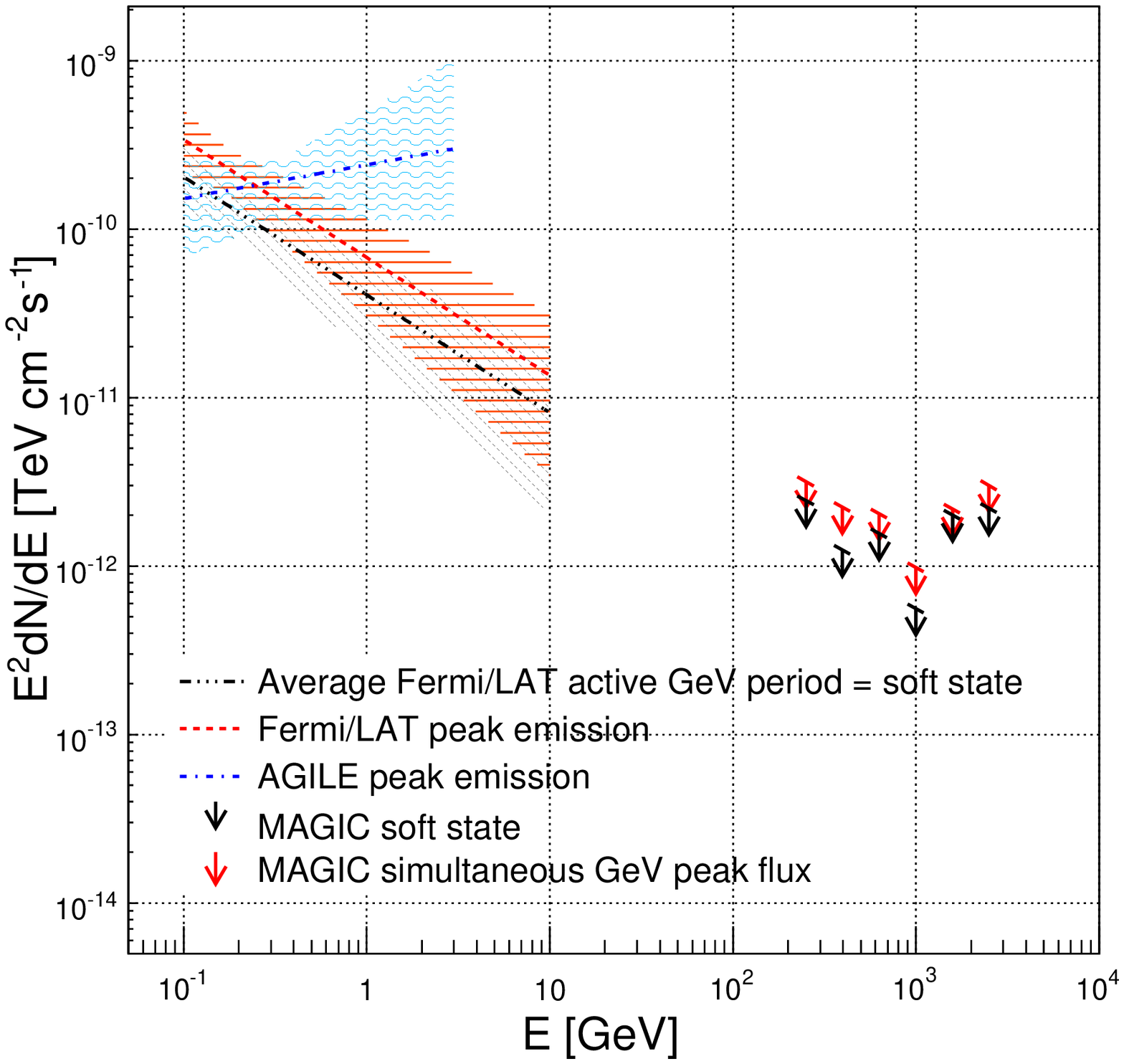}     
\caption{Left: VHE gamma-ray and X-ray light curves of LS~I~+61~303 during
the multiwavelength campaign of 2007 September. Right: Cygnus X-3 SED in the
high-energy and VHE bands. The lines indicate the power-law spectra derived from Fermi/LAT and AGILE integral
fluxes and photon indices, where the corresponding errors were taken into
account and are shown in shadowed areas. The arrows display the \pcnt{95} CL
MAGIC differential flux upper limits.}
\label{fig:ls1_x3}
\end{figure}

Another kind of gamma-ray emitter are binary systems. One of them is the X-ray
binary LS~I~+61~303, for which MAGIC conducted a multiwavelength campaign with
XMM-Newton and Swift during \pcnt{60} of an orbit in 2007 September. A
simultaneous outburst at X-ray and VHE bands was detected, as shown in
Figure~\ref{fig:ls1_x3} \cite{ref:ls1}. The simultaneity,
and the extracted X-ray/VHE flux ratio, suggested that, for this outburst, the
X-rays are the result of synchrotron radiation of the same electrons that
produce VHE emission as a result of inverse Compton scattering of stellar
photons.

Cygnus X-3 is a microquasar consisting of an accreting compact object orbiting
around a Wolf-Rayet star. It has been detected in gamma-rays above 100 MeV and
many models also predict a VHE emission when the source displays relativistic persistent jets
or transient ejections. MAGIC observed Cygnus X-3 for about 70 hours between
2006 and 2009 in different X-ray/radio spectral states and also
during a period of enhanced gamma-ray emission \cite{ref:x3}. No signal was found, and
the most stringent upper limits to date were extracted for different spectral
states, orbital phases, energies, and dates. Figure~\ref{fig:ls1_x3} shows the
SED for different spectral states.


\section{Upgrade plans}

In 2011, the MAGIC telescopes will be upgraded. The goal is to make the two
telescopes more similar, and thus achieve a more homogeneous
exposure. The
main targets of the upgrade includes a complete replacement of the MAGIC-I
camera, equipping it with an equally large number of pixels (1039) and an equally
large trigger area. Besides that, both telescopes will get a new readout readout based on Domino-4
ring samplers\footnote{http://drs.web.psi.ch} and new low-energy specialized
SUM-trigger systems.


{\scriptsize \section{Acknowledgments}


We would like to thank the Instituto de Astrof\'{\i}sica de
Canarias for the excellent working conditions at the
Observatorio del Roque de los Muchachos in La Palma.
The support of the German BMBF and MPG, the Italian INFN, 
the Swiss National Fund SNF, and the Spanish MICINN is 
gratefully acknowledged. This work was also supported by 
the Marie Curie program, by the CPAN CSD2007-00042 and MultiDark
CSD2009-00064 projects of the Spanish Consolider-Ingenio 2010
programme, by grant DO02-353 of the Bulgarian NSF, by grant 127740 of 
the Academy of Finland, by the YIP of the Helmholtz Gemeinschaft, 
by the DFG Cluster of Excellence ``Origin and Structure of the 
Universe'', and by the Polish MNiSzW Grant N N203 390834.}

%


\end{document}